# Trapping of 27 bp - 8 kbp DNA and immobilization of thiol-modified DNA using dielectrophoresis

**Sampo Tuukkanen[1], Anton Kuzyk[1], J Jussi Toppari[1],**
**Hannu Häkkinen[2], Vesa P Hytönen[3,4], Einari Niskanen[3],**
**Marcus Rinkiö[1] and Päivi Törmä[1]**

[1.] Nanoscience Center, Department of Physics, PO Box 35 (YN), FIN-40014 University of Jyväskylä, Finland
[2.] Nanoscience Center, Department of Chemistry, PO Box 35 (YN), FIN-40014 University of Jyväskylä, Finland
[3.] Nanoscience Center, Department of Biological and Environmental Science, PO Box 35 (YAB), FIN-40014 University of Jyväskylä, Finland
[4.] Present address: Department of Materials, ETH Zürich, CH-8093 Zürich, Switzerland.

**E-mail:** sampo.tuukkanen@phys.jyu.fi

**Abstract.** Dielectrophoretic trapping of six different DNA fragments, sizes varying from the 27 to 8416 bp, has been studied using confocal microscopy. The effect of the DNA length and the size of the constriction between nanoscale fingertip electrodes on the trapping efficiency have been investigated. Using finite element method simulations in conjunction with the analysis of the experimental data, the polarizabilities of the different size DNA fragments have been calculated for different frequencies. Also the immobilization of trapped hexanethiol- and DTPA-modified 140 nm long DNA to the end of gold nanoelectrodes was experimentally quantified and the observations were supported by density functional theory calculations.

**Keywords:** DNA, dielectrophoresis, polarizability, hexanethiol, DTPA, immobilization, confocal microscopy.

**PACS:** 87.15.-v, 82.37.Rs

## 1. Introduction

Due to its exceptional self-assembly properties DNA has already proven its applicability in future molecular electronic applications, e.g., as a scaffold [1] or even as a charge carrier. Whether fabricating devices constructed of single molecules or larger self-assembled entities the final step still includes the embedding of the device into the rest of the circuit, and thus, precisely controlled positioning of DNA at single molecule level is required.

Dielectrophoresis (DEP) means the induced motion of polarizable, neutral or charged, particle in the inhomogeneous electric field [2]. In the case of 'positive' ('negative') DEP, polarized objects are moving towards the electric field maximum (minimum). 'Negative' DEP is the effect of the surrounding medium polarizing more than the objects themselves. The dielectric force is determined by $F_{DEP} = 1/2 \cdot \alpha \cdot \nabla(E^2)$, where (in the case of ac field) $\alpha$ is in-phase component of the polarizability of the object (in the certain medium) and $E$ is the root mean square value of the electric field (assuming sinusoidal time dependence). The dielectrophoretic force compared with the Brownian motion of the particle together with the viscosity of the medium determines the efficiency of trapping.

DEP has been demonstrated to be useful in many different fields [3, 4]. In the micrometer scale, DEP has been widely used as an active, non-destructive manipulation method for trapping biological objects like eukaryotic [5, 6] and prokaryotic (bacterial) cells [7], and DNA of different lengths [8], i.e. mostly 16 µm long λ-DNA [9-16].

For nanometer-scale objects, Brownian motion poses a greater challenge. In nanoscale, DEP has been demonstrated for latex beads [17], viruses [17, 18], nanoparticles [19], and proteins [16, 20, 21]. There are only a few successful demonstrations for trapping nanoscale DNA molecules: a) circular 2.7 kbp pUC18 plasmid (0.9 µm in perimeter) [22], b) 368 bp (~125 nm) double-stranded DNA (dsDNA) and 137 b (base) (~76 nm) single-stranded DNA (ssDNA) [23], and c) 40 bp (~14 nm) dsDNA, 40 b (~22 nm) and 1 kb (~560 nm) ssDNA [24].

In our previous work, we have demonstrated that single DNA molecules of the length scale of 100 nm can be trapped and immobilized using DEP with metal electrodes [25]. In a subsequent work [26], we showed that the efficiency of the trapping process can be improved by the use of a carbon nanotube as one electrode, and that the DEP experiments can be utilized to obtain information about the polarizability of DNA as a function of frequency. In both [25] and [26], DEP was only demonstrated for a small set of parameters, i.e. the frequency and field dependence of DEP was not systematically studied and thereby the process was not optimized. Furthermore, the dependence of the DEP process and DNA polarizability on the length of the DNA was not studied. In the present paper we carry out a systematic study of the DEP method and DNA polarizability where we vary all the relevant parameters: the frequency (0.2 - 10 MHz), the voltage of the applied signal (0.35 - 3.5 $V_{rms}$), and the length of the DNA (27 bp – 8461 bp). The aim of such studies is twofold: to find an optimized set of parameters for the DEP method, and to gain insight into the origin of DNA polarizability. In addition, since we aim to present a complete method for not only trapping but also immobilizing DNA, we also study, both experimentally and theoretically, the effect of the thiol-modification of the DNA on the efficiency of immobilization.

In this paper, we systematically analyze the DEP of dsDNA of different lengths, using alternating (AC) electric fields of different frequencies and voltages applied on fingertip type metal electrodes. The analysis is performed *in situ* under confocal microscope. These results give directly information about the efficiency of the trapping process as a function of the parameters varied. Furthermore, information about DNA polarizability can be inferred from this data, but this requires finding the connection between the applied voltage and the field profile it creates, which we obtain by finite-element method (FEM) simulations of the electric field in the used nanoelectrode structure. The polarizability is determined as a function of both the frequency of the applied voltage and the length of the DNA. Finally, we compare the immobilization of DNA using two different types of thiol-modifications, namely hexanethiol and DTPA (dithiol-phosphoramidite), to immobilization of unmodified DNA. DTPA has not previously been reported to be used as a DNA-linker. We explain the qualitative behaviour observed in the immobilization experiments by binding energies obtained using density functional theory (DFT) calculations.

The key findings of the present paper are the following. Higher electric fields are required for trapping smaller DNA. This is consistent with the finding that DNA polarizability decreases with its length. An interesting result is, however, that DNA polarizability per base pair is bigger for smaller molecules than for longer ones. This sheds new light on the role of the shape of DNA and the counter-ion cloud in DNA polarizability. The frequency dependence of the DEP process and of DNA polarizability was found to be rather weak. In general, more DNA is trapped with lower frequencies, on the other hand for higher frequencies the DNA is better localized in the desired point. This competition between efficiency and accuracy results into the optimum frequency which was found to be ~1 MHz. The hexanethiol modification was found to be better for immobilization than the DTPA, and this finding is qualitatively consistent with the binding energies obtained by DFT.

The contents of this paper are organized as follows: The fabrication of materials and electrodes are described in section 2, while the experimental methods and data analysis are presented in section 3. In section 4, we discuss the DEP theory and present the FEM simulation methods. The results of DEP of different size DNA fragments, including frequency and length dependence as well as the determined polarizabilities, are given in section 5, and DNA immobilization results in section 6. In section 7, we summarize the results.

**2.     Materials and methods**

*2.1.     DNA fragments*

Double-stranded DNA fragments (see Table 1) with varying lengths (27 – 8461 bp) were fabricated by three different methods: 1) Annealing of the synthetic oligonucleotides. 2) PCR (Polymerase Chain Reaction). 3) Restriction enzyme digestion of the plasmids multiplied in bacteria. The concentrations of final products were measured spectrophotometrically.

27 bp fragment were generated by mixing equal amounts of the complementary oligonucleotides (TAGC, Copenhagen, Denmark) *Primer1* and *Primer2* in 6.5 mM Hepes (N-2-Hydroxyethylpiperazine-N'-2-ethanesulfonic acid) buffer (pH 7.0 adjusted with NaOH).

145 and 444 bp DNA fragments were produced by the PCR reaction using TAQ polymerase (Fermentas) with the oligonucleotides *Primer3* and *Primer4* or *Primer4* and *Primer5*, respectively, as primers in the PCR reaction. The PCR was followed by a purification by 1 % agarose gel electrophoresis and an isolation with GFX™ PCR, DNA and Gel Band Purification kit (Amersham Biosciences). Chicken avidin complementary DNA in pFastBac1-plasmid (Invitrogen) was used as a template in the PCR syntheses [27].

1065 bp DNA fragment was generated by digesting the pBVboostFG plasmid [28] using BglI & SpeI restriction enzymes. 5141 bp fragment was produced by linearizing the pFastBac1 plasmid (Invitrogen) using HindIII enzyme. Finally, 8461 bp fragments were generated by linearizing pBVboostFG using ApaI enzyme. The restriction enzymes were purchased from Promega. The fragments were purified as described above. The plasmids were produced by cultivating transformed *E. coli* JM109 cell line (Stratagene) at 37 °C in the suspension and isolating the plasmids from the overnight cultures by using the plasmid purification kit (Macherey-Nagel, Düren, Germany).

**Table 1.** The oligonucleotides used in fabrication of DNA fragments.

| Name | Sequence |
|---|---|
| *Primer1* | 5'-GGT GAA TTC GCC GGC ACC TAC ATC ACA-3' |
| *Primer2* | 5'-TGT GAT GTA GGT GCC GGC GAA TTC ACC-3' |
| *Primer3* | 5'-CCC GAT GGT CAT GTT GGC GCC CAG ATC GTT GGT-3' |
| *Primer4* | 5'-CTG CTA GAT CTA TGG TGC ACG CAA CCT CCC C-3' |
| *Primer5* | 5'-GAG TGA AGA TGA TGA TGC CGA CC-3' |
| *Primer6* | 5'–HS–(CH$_2$)$_6$–GCC AGA AAG TGC TCG CTG AC–3' |
| *Primer7* | 5'–HS–(CH$_2$)$_6$–TTC TCG ACA AGC TTT GCG GG–3' |
| *Primer8* | 5'-DTPA-GCC AGA AAG TGC TCG CTG ACT G-3' |
| *Primer9* | 5'-DTPA-CTT CTC GAC AAG CTT TGC GGG-3' |

*2.2.     Thiol-modified DNA fragments*

We used two different types of linkers: a hexanethiol and DTPA (dithiol-phosphoramidite). Double-stranded DNA containing a modification group in both ends was obtained by using 5'–modified oligonucleotides as primers in PCR. Oligonucleotide *Primer6* (purchased from Synthegen, Houston, Texas) was used as a forward primer and *Primer7* as a reverse primer for hexanethiol modified 414 bp DNA (from now on called C6-DNA). Oligonucleotide *Primer8* (purchased from TAG Copenhagen A/S) was used as a forward primer and *Primer9* as a reverse primer for DTPA modified 415 bp DNA (from now on called DTPA-DNA). The PCR was done as above in section 2.1.

*2.3.     DNA solutions for the experiments*

Prior to use, DNA fragments were diluted into Hepes/NaOH buffer, i.e., 3 mM Hepes and 1 mM NaOH, which yield pH 6.9 and conductivity 20 μS/cm (conductivity meter, model CDM3, Radiometer, Copenhagen), and the obtained solution was stored in a refrigerator in small aliquots. Low conductivity was necessary to prevent the excess Joule heating of the buffer [29], to reduce the oxidation-reduction reactions at the electrode-solution interface, and to obtain high polarization of the DNA relative to the polarization of the buffer (lower conductivity induces lower polarization of the buffer and also the thickening of the Debye layer around the DNA, i.e., counter-ion cloud) [22]. The DNA was labelled with the dsDNA specific fluorescent label PicoGreen (Molecular Probes, Eugene, USA) diluted 1:200 into Hepes/NaOH buffer. The final concentrations of the fragments were chosen so that the concentration of the nucleotides remained the same in all the cases (17 μM nucleotides). Since the fluorescent dye molecules attach approximately uniformly along the helix [30], this ideally results the same amount of fluorescence in the solutions of DNA molecules of different length. The final concentration of PicoGreen was 1.6 µM yielding a dye to base pair ratio of 1:5 [31].

For the immobilization experiments, 10 nM solutions of C6-DNA and DTPA-DNA diluted into Hepes/NaOH buffer were used. In addition, 0.5 mM TCEP-HCl (Tris(2-Carboxyethyl) Phosphine and Hydrochloride) or alternatively 2 mM $NaBH_4$ (Sodium borohydride) was added to the solutions about one hour before the DEP experiment as a reduction agent to break sulphur-sulphur bonds between separate DNA molecules and to make thiol-groups more reactive. Prior to the experiments 1.6 µM PicoGreen was added (as above). It should be noted that the presence or absence of the reduction agent did not significantly affect on the amounts of immobilized thiol-modified DNA when bigger amounts were trapped. However, the use of reduction agent is essential to avoid a possible attachment of multimers when aiming at the immobilization of individual molecules [25].

*2.4. Finger-tip electrodes*
The fingertip type electrodes were composed of narrow (about 100 nm wide) wires with about 100 nm separation in between. They were fabricated on a slightly boron-doped (100)-silicon substrate with a thermally grown $SiO_2$ at the top. Polymethylmethacrylate (Microchem C2 PMMA) resist was spin-coated and patterned by the electron beam writer (Raith eLine, equipped with Elphy Quantum 4.0 -lithography software). The evaporation of metal took place in the ultrahigh vacuum (UHV) chamber. The thickness of the evaporated gold layer was 15 nm, under which 2-5 nm of titanium was used to improve the adhesion of gold. The samples were cleaned using a short flash of oxygen plasma (Oxford Plasmalab 80 Plus RIE, parameters: 100 sccm O2-flow, 50 W RF power and 1 min time) before the confocal microscope experiment to clean the gold from organic contaminants which would prevent thiols to chemically bond to the gold surface.

*2.5. AFM imaging*
Atomic force microscope (AFM) (Veeco Dimension 3100) was used in the characterization of the finger-tip electrode samples. The AFM was operated in tapping-mode using silicon probes (Veeco MPP-11100), which have the resonance frequency of 300 kHz and the spring constant of 40 N/m.

## 3. Experimental

*3.1. DEP under the confocal microscope*
Dielectrophoresis experiments under the confocal microscope (Zeiss Axiovert LSM510, Zeiss "Fluar" 40x/1.3 Oil objective) were performed by (fluorescent) imaging a $10\times10$ µm$^2$ square area around the gap between the fingertip electrodes during the trapping of DNA using an AC signal (Agilent 33120A waveform generator) applied to the electrodes. The method (previously used in ref [26]) is represented in figure 1. For imaging, argon laser (488 nm) with power of 0.45 mW was used, and tested in advance not to cause too much bleaching of the fluorescent dye (See section 3.3). Fluorescence data was collected simultaneously from two channels: the fluorescence channel (equipped with 505 nm high pass), which corresponds to the amount of DNA, and the reflection channel (equipped with 475-525 nm band pass), which shows the location of the electrodes. Dielectrophoresis movies were obtained by capturing two 128×128 pixel frames per second. In the beginning of each DEP movie, the voltage is kept off for 10 s after which the sinusoidal AC signal is turned on (to a certain starting voltage value). The voltage was raised by 0.2 $V_{p-p}$ (0.07 $V_{rms}$) after each 20 s period until the final voltage value was reached. The voltage was then turned off but the data collection was still continued for 20 seconds to see how DNA diffuses away from the gap.

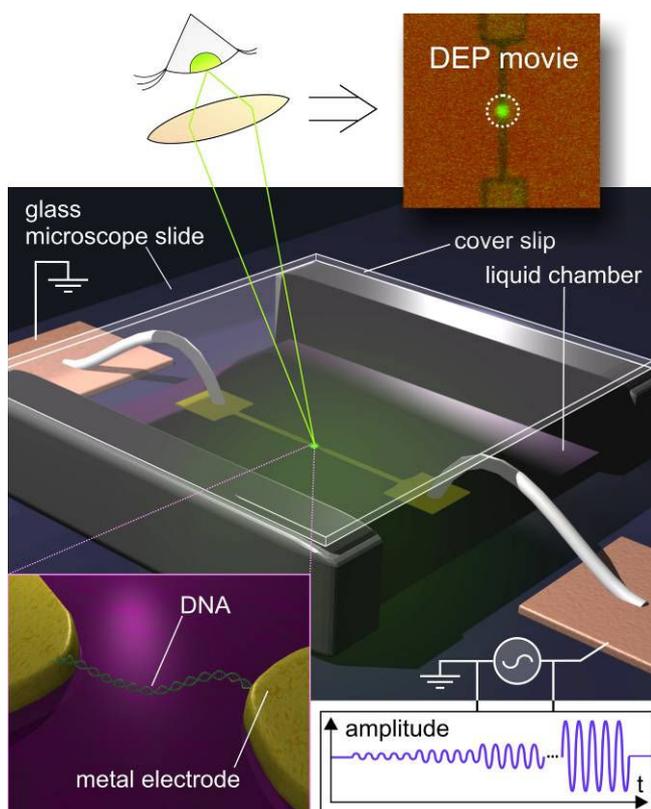

**Figure 1**. A schematic view of the experimental setup used in the DNA DEP experiments under the confocal microscope. In the middle, there is the nanoelectrode structure, and the chamber containing the DNA solution at top of it. Closer view of the DEP trap is presented in the lower left image. An example of a DEP movie obtained from the captured confocal microscope images is shown in the upper right image. The lower right image represents the sinusoidal ac signal taken from the waveform generator.

*3.2. Fluorescence data analysis*

For the quantitative analysis, the amount of DNA collected in the gap was obtained from each frame of the fluorescence movie by determining the mean fluorescence intensity inside the circle shaped (diameter of 1.6 μm, see upper right inset in figure 2) area in the gap between the fingertip electrodes subtracted by the mean intensity of the background fluorescence (measured from the circle shaped area on substrate surface a few μm distance from the gap). To optimize the measurement for obtaining accurate information about the voltage and frequency dependence of the trapping process, especially to exactly determine the minimum voltage $V_{min}$ for which the trapping of DNA begins, the detector sensitivity was maximized by fine-tuning the detector gain and the amplification offset according to the fluorescence background of each sample. This makes (for technical reasons) the absolute fluorescence values not exactly comparable between different samples, but it was needed to be able to distinguish from the background the very small changes in the fluorescence due to DNA. Even when the absolute intensities are not directly available, we obtain consistent information about the frequency and voltage dependence of the trapping process, which our conclusions are based on. For the plotting of the fluorescence data, the obtained fluorescence curves were (in most cases) normalized by setting the maximum fluorescence intensity observed for each sample to unity. The data analysis of the fluorescence movies was done using the confocal microscope software (LSM 510), Matlab 6.1 and Origin 7.5.

*3.3. Electrochemical effects and dye bleaching*

When an electric field is acting in a conducting media, the media heating occurs with a power $W = \rho E^2$ per unit volume [29]. This can cause denaturation of DNA driving it to a single-stranded conformation. The buffer conductance was measured to be about $\rho = 20$ μS/cm and the approximate electric field in the gap is about $10^7$ V/m yielding the power generation $10^{12}$ J/m³s. For a cubic volume of (100 nm)³ (which is about the volume of the gap region) this gives 1 nW, which is very low power and the induced heating is fractions of a Kelvin at its maximum [32]. Also the fact that the volume where DEP trapping occurs is small yields large surface to volume ration, thus making heat dissipation very efficient. This further means that the heating of the medium is not a problem in our case.

When the trapping voltage is large enough the role of hydrodynamic effects increases and can induce convectional flows which disturb the trapping process [21]. In the experiments, we have sometimes observed this kind of disturbance in the voltage-intensity curves when relatively high voltages were used.

The bleaching tests for PicoGreen fluorescent dye labelled C6-DNA were performed with the confocal microscope using the same parameters and settings that were used in capturing of the DEP movies. The bleaching tests were done using PicoGreen labelled C6-DNA immobilized to gold electrode. The fluorescent dyes attached to the DNA remained functional for quite a long time (at least tens of minutes) during the laser excitation, which makes PicoGreen suitable for this kind of DEP studies. It was estimated from the fluorescence intensity vs. time curves that the highest bleaching rate was about 10 % per minute.

**4.     Obtaining the DNA polarizability from the DEP results**

Since dielectrophoresis is dependent on the polarizability of the object and on the applied electric field, our *in situ* observations of DNA DEP contain indirect information about DNA polarizability. To infer this information, we first need to obtain a connection between the voltage applied to the nanoelectrodes and the electric field that it produces. This is done by finite element method (FEM) simulations as explained below in section 4.1. Second, we need to define a connection between the trapping observations and the trapping potential that the DNA actually feels, this is done in section 4.2.

*4.1.     Finite element method simulations*

3D Poisson's equation was solved using FEM solver (COMSOL Multiphysics 3.2a) to determine the 3D electric field $\vec{E}(\vec{r})$ created by the fingertip electrode structure inside a (8 μm$^3$) cubic volume. For the frequencies used in our experiments, a dilute Hepes/NaOH buffer can be approximated as a homogenous medium. Homogenous media are assumed in the used FEM method. For 1 μm thick buffer layer, we used the permittivity of water, $\varepsilon_m = 80$, which is a good approximation for the dilute buffer used. Under the electrodes, there was 300 nm layer of SiO$_2$ ($\varepsilon_r = 3.7$) and 700 nm layer of silicon ($\varepsilon_r = 11.7$). The values for the relative permittivity were chosen to correspond to the situation of using 1 MHz electric field.

The simulations were performed for the different values of the separation between the fingertip electrodes, i.e., 100 and 200 nm, and for the dc voltage applied between the electrodes (See insets in Fig. 2). This dc value can be interpreted as a RMS-value in the case of ac voltages. One can get the DEP force, $F_{DEP} = 1/2 \cdot \alpha \cdot \nabla(E^2)$, by multiplying the effective polarizability of the molecule, $\alpha$, by the gradient of electric field square, $\nabla(E^2)$, obtained from the FEM simulations.

*4.2.     Limit for trapping*

Due to the polarization of the positively charged counter-ion cloud [11, 22, 33], DNA in aquatic solution is highly polarisable thus yielding a high DEP force. This trapping force is competing against the thermal drag force [2] (or threshold force, which is defined from the diffusion path during the experiment [29]) of a certain size object. This gives a theoretical minimum voltage for successful trapping using a certain electrode structure. The thermal drag force is roughly $F_{Th} = k_B T / 2r$ [2], in the case of spherical particles with radius *r*, but in the case of DNA it can not be applied as such. Very long DNA molecules, which have a contour length substantially larger than several hundreds of base pairs, are of globular shape in their native state, but during the DEP, they have been shown to elongate near to their contour length [9, 10, 13, 15]. This kind of thin, long objects have an enhanced polarizability [34]. Therefore long DNA molecules can be considered as micron-scale objects and they can be easily manipulated using DEP. Yet, to estimate the thermal drag force one can still use the radius of the ball as a first approximation. For example, if 12 kbp plasmid DNA is packed into a tight sphere it would have a radius of about 19 nm [33], which yields a thermal drag force of ~100 fN.

In the case of DNA fragments, which are smaller or about the same size than the persistence length of DNA, which is typically ~150 bp (~50 nm) [35], the molecule behaves more like a rigid rod of 2 nm axial diameter, and its Brownian motion is more difficult to define. However, from a tightly packed approximation we get an estimation for an upper limit of the thermal drag force. If 100 bp DNA is at tightly packed sphere it has a radius of 4 nm, which yields the maximum thermal drag force of ~500 fN. Thus, one can estimate that the thermal drag force for long and short DNA fragments are approximately the same, i.e., a few hundreds of fN. However, it should be noted that these values are upper bounds and dsDNA usually never appears as tightly packed as approximated above.

Instead of the drag force we can also use the thermal energy $U_{Th} = \frac{3}{2}k_BT$ associated with Brownian motion. The 'DEP potential', i.e., potential energy during DEP trapping, is $U_{DEP}(\vec{r},\omega) = -\frac{1}{2}\alpha E^2$, where the effective polarizability $\alpha$ depends on the frequency $\omega$ of the applied signal and on the properties of the molecule [36]. The use of an ac field averages the electrophoretic forces acting on the negatively charged DNA to zero and we obtain for the total potential energy $U_{tot} = U_{Th} + U_{DEP} = \frac{3}{2}k_BT - \frac{1}{2}\alpha E^2$, which has a minimum at the point of highest electric field. The DEP potential energy is plotted in figure 2 as a function of the perpendicular distance from the gap for different applied voltages and gap sizes. The probability density, and thus the density of DNA molecules at a certain position, $\vec{x}$, of the DEP potential field is given by an exponential function $\rho(\vec{x}) \propto \exp[-U_{DEP}(\vec{x})/k_BT]$ [37]. Due to a limited accuracy of the experimental data, in our case it is sufficient to approximate this as a step function/box potential. That is, in the following we will simply base our analysis on equating the two competing energies, $U_{DEP}(\vec{x}) \approx k_BT$. This yields a condition $U_{tot} \leq 0$ for successful trapping.

By determining experimentally the minimum electric field (on the edge of the trapping area, See section 5.4) needed to trap a certain size molecule one can calculate its polarizability $\alpha$. Further, since in the experiments the electric field was generated by the voltage applied to the nanoelectrodes, the task equates to determining of the minimum voltage $V_{min}$ needed for trapping. This corresponds to the point where the trapping begins while increasing the voltage, i.e., observed fluorescence in the DEP trap exceeds the background noise level and starts to rise as $\propto V^2$. The $V^2$ dependency is physically motivated by the DEP force, $F_{DEP} = 1/2 \cdot \alpha \cdot \nabla(E^2)$, and has also been observed experimentally [38]. However, since the measured fluorescence, i.e., the amount of the trapped DNA, is a statistical property, the change from a constant value $I_0 + A \cdot V_{min}^2$, which is very small, to the $V^2$ dependency is not abrupt but smeared. This has been taken into account by determining $V_{min}$ via fitting the fluorescence intensity to the function $I = I_0 + A \cdot (V^b + V_{min}^b)^{2/b}$ which produces the correct dependencies above and below $V_{min}$, but in addition includes the parameter $b$ determining the rate of the change (the best fit was found using $b = 40$, which corresponds to a slightly smeared change). From the voltage $V_{min}$, obtained by fitting the data to the function, one can then calculate the corresponding electric field using the FEM simulations (See section 4.1). The polarizability can also be obtained using an experimentally determined minimum force and the thermal drag force, but due to the more ambiguous description of the thermal drag force the potentials were used in calculations of the section 5.3.

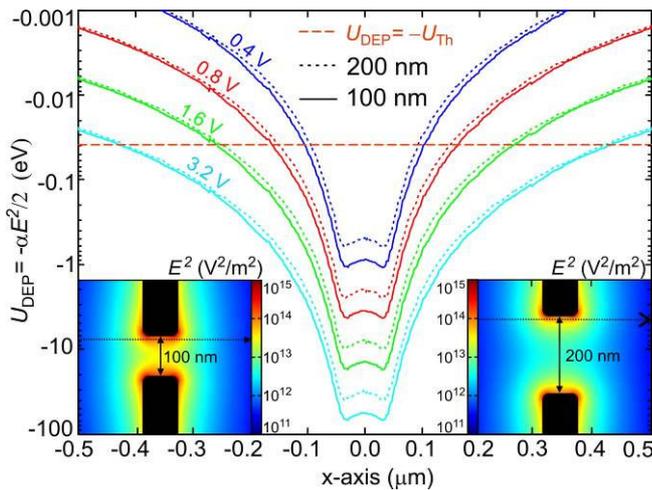

**Figure 2.** The 'DEP potential energy' represented as a function of the perpendicular distance from the electrodes along a dotted line shown in the insets, 5 nm apart from the electrode end (to avoid the equipotential surface of the electrode). For this illustration of $U_{DEP}$ we used the total polarizability $\alpha = 10^{-34}$ Fm$^2$/bp·100 bp $= 10^{-32}$ Fm$^2$, calculated for 100 bp DNA (as an example) using a typical value found from the literature [22, 26, 33]. The results are given for the fingertip electrode separations of 100 and 200 nm and for the dc voltages 0.4, 0.6, 1.2 and 3.2 V. The horizontal dotted line represents the level where the thermal energy at T = 300 K cancels the DEP potential energy. Used dc voltage values correspond to RMS-values in the case of ac signals. The electric field square, $E^2$, obtained from the simulations for the voltage 0.8 V and the electrode separations of 100 nm and 200 nm is shown in the insets.

## 5. DEP of DNA fragments

*5.1. Trapping efficiency vs. gap size*

The fingertip electrode samples with different sizes of the gap, i.e., 80 and 130 nm, were used to trap C6-DNA and DTPA-DNA using the trapping voltages from 0.7 to 1.4 $V_{rms}$ (See figure 3). From figure 3, one can observe that the trapping starts always after the same threshold voltage, $V_{min}$, independently on the size of the gap. After this voltage the raise in intensity has the $V^2$ dependence, as expected. This dependency was proven, and the value of $V_{min}$ was determined, for each curve via fitting as explained in section 4.2.

This result states that the trapping efficiency has no clear dependency on the gap size, which seems first counterintuitive to the description of the DEP force. However, from the figure 2 one can see that the region where the electric field clearly depends on the gap size is small, i.e., ~100 nm around the middle point, and outside of that the DEP potential depends mainly on the absolute voltage applied between the electrodes. Since the minimum size of the fluorescence spot that could be unambiguously resolved in our experiments was ~1 μm due to, e.g., the spreading of DNA spot by the electrophoretic effects [23] and the resolution limitations such as the steps in the trapping voltage, the "trapping region" is large compared to that region. This explains the lack of the gap dependency in the data of figure 3. This also indicates that our trapping region is indeed larger than the gap size and the observed ~1 μm spot is not optically limited (estimated optical resolution ~200 nm).

Note that a larger change in the gap size, i.e., change comparable or larger than the trapping region ~1 μm, would be visible in the obtained intensity curves and thus interpreted as a change in the trapping efficiency. However, in the earlier work by the authors [26], where a similar fingertip electrode was compared to a significantly narrower carbon nanotube electrode, it was shown that the high electric field gradients in the end of carbon nanotube made it a much more efficient 'DEP trap' even if the electrode separation is large. Thus, one can deduce that the shape of the electrode has more effect to the trapping efficiency than small changes in the gap size.

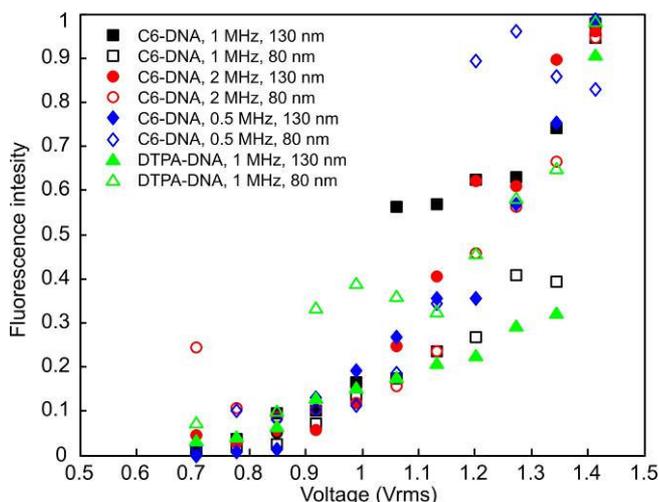

**Figure 3.** The fluorescence intensity of trapped C6-DNA or DTPA-DNA is plotted as a function of the voltage $V_{rms}$ using samples with different size gaps (80 or 130 nm).

*5.2. Trapping efficiency vs. DNA length*

Here we discuss the trapping of different size DNA fragments (27 – 8461 bp) as a function of the voltage using different frequencies (0.2, 0.5, 1, 2, 5 and 10 MHz). Measured fluorescence intensity in the DEP trap as a function of the trapping voltage for the different fragments, using 1 MHz frequency, is represented in figure 4. The fluorescence intensity corresponds to the amount of nucleotides, because the dyes are supposed to attach uniformly along the DNA helix. From the fluorescence curves one can easily see that when the contour length of the molecule is short, more voltage is needed for the trapping to start, i.e., the intensity starts to rise at higher voltage and fitting, as described in chapter 4.2, yields higher values for $V_{min}$, as shown in the inset of the figure 4. This is due to two issues. For smaller molecules (1) the Brownian motion is higher and (2) the polarization of the molecule is smaller which results in the smaller DEP force. However, DEP works also in the case of very short DNA fragments (less than ~50 bp). This is due to the polarization of DNA being mainly caused by the counter-ion cloud, which has a certain minimum thickness, i.e., the Debye layer which is ~10 nm in the case of NaOH/Hepes buffer [22]. For instance, in the case of 27 bp, which is about ~9 nm long rod-like object, the counter-ion cloud makes its effective length substantially longer. This enhances the polarizability of small DNA fragments and increases also the DEP force.

The contour length of the largest DNA fragment (8461 bp) we used was about 2.8 μm. However, in its native state it is a randomly coiled globular ball about the radius of even as small as 17 nm [33]. In many studies, it has been observed that long DNA molecules are straightened (or stretched) during DEP [9, 10, 13, 15]. In this work one could not observe this kind of straightening of DNA because the fingertip electrode separation was only 100 nm and the resolution limit of confocal microscope prevents one to distinguish the shape of objects of ~100 nm size.

An average electric field (voltage divided by the gap size) we used for trapping of DNA was from $10^7$ to $9 \cdot 10^7$ V/m. In earlier studies, in the case of electrode constrictions of 1 – 10 μm and DNA molecules larger than 10 kbp, the field strength between $10^5$ and $10^6$ $V_{rms}$/m was enough for trapping [15, 22, 23, 33]. To realize trapping in the case of smaller DNA fragments, the field strength higher than $10^6$ $V_{rms}$/m was needed [8, 22, 23]. However, when the size of a constriction is suppressed to a nanoscale even higher electric field strength has been used, e.g., in the DEP of a protein of about the same mass as 400 bp DNA by using the gap of 500 nm, the electric field of ~$2 \cdot 10^7$ $V_{rms}$/m has been used [21]. In our experiments the gap was only ~100 nm, creating the highest field inside a region of similar size (See Fig. 2). To obtain the trapping region comparable to the size of the observable spot (as large as ~1 μm), the voltage had to be further increased yielding even higher average field strengths.

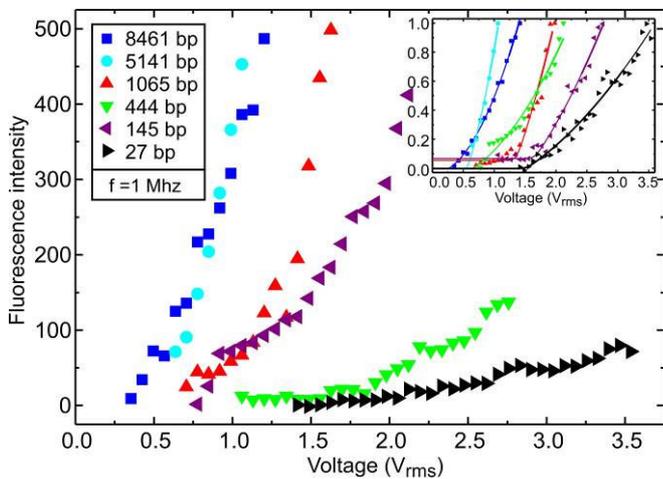

**Figure 4.** The trapped amounts (the fluorescence in arbitrary units) of different size DNA fragments as a function of the trapping voltage using 1 MHz frequency. The relative fluorescence amounts are nearly comparable to each other, because almost the same capturing settings have been used. In the inset, the normalized fluorescence curves (maximum fluorescence point is set to unity) and fits to the function $I = I_0 + A \cdot (V^b + V_{min}^b)^{2/b}$ (see section 4.2.) are represented.

### 5.3. *Effects of the frequency on trapping*

In the case of all DNA fragments that were used, it was observed that the higher frequencies were used the higher voltages had to be applied to realize trapping. This can be understood so that when the trapping frequency is higher the molecule has less time to polarize yielding the lower polarization, as shown in figure 5, and thus lower DEP force (see chapter 5.4). For the frequencies higher than 10 MHz the trapped amount of DNA decreased so much that no noticeable fluorescence intensity change was observed. This may be due to the frequency dependency of DNA DEP or to the signal losses in the narrow (~100 nm) electrodes, which are not optimized for the high frequency signals. Although the smallest observable fluorescence spot when the trapping begins (at voltage $V_{min}$, see section 5.4) is nearly the same for all frequencies, for higher voltages the spot size depends on the frequency: too low frequency results to the spreading of the spot by the electrophoretic force, that is, the occupancy of localization of DNA is poor. This was also observed earlier by the authors [25].

### 5.4. *Polarizability of DNA fragments*

From the fluorescence-voltage curves (see figure 4) one can obtain frequency and voltage dependent information about the DEP of different size DNA fragments, e.g., the polarizability. To calculate polarizabilities of the DNA fragments from the experimental data, we first determined the minimum voltage $V_{min}$ needed for trapping. Next, we use the information obtained from the simulations to find out the electric field strength $E_{min}$ on the edge of the fluorescence spot, i.e., on the edge of the DEP trap, when the corresponding minimum voltage $V_{min}$ is applied to the electrodes. The radius of the smallest observable fluorescence spot (when the trapping begins) was approximately $r = (0.5 \pm 0.1)$ μm for all the frequencies (Note that even the fluorescence spot sizes have a frequency dependence, it cannot be resolved in the beginning of trapping, but only later with larger trapped amounts). The specifying of the smallest resolved fluorescence spot to ~1 μm is not very accurate because it is partially limited by the optical resolution of confocal microscope (i.e. ~200 nm) and it can give a small systematic error to the obtained polarizability values. For the polarizability, $E_{min}$ was read from the simulated data at the distance $r$ from the end of the fingertip electrodes, perpendicular to them, in the plane 10 nm above and parallel to the substrate surface. Now, by setting the total potential energy to zero ($U_{tot} = U_{DEP} + U_{Th} = 0$) on the edge of the fluorescence spot, we obtain the polarizability as $\alpha = 3k_B T / E_{min}^2$.

The calculated polarizabilities per base pair (the polarizability divided by the molecule length in base pairs) are shown as a function of frequency in figure 5 and as a function of DNA length in figure 6. From figures 5 and 6 one can see that polarizability per bp is larger in the case of short DNA fragments, which is an indication of polarization of the counter-ion cloud (discussed before in section 5.2) [11, 22, 33]. The linear fits to the polarizability values of each fragment in figure 5 show that polarizability slightly decreases with the increase of the frequency. An exception to this is 1 MHz frequency, which seems to give the highest polarizability values for almost in the case of all fragments. It has been shown earlier, that 1 MHz frequency works well for DNA DEP [13, 14, 16, 20, 22, 25]. The total polarizability per molecule (inset of figure 6) was observed to increase as a function of the molecule length as expected.

In the case of long DNA molecules, i.e., substantially longer than the persistence length of DNA (~150 bp) [35], the dependence of the polarizability on the DNA length can be understood via Manning's model [39]. In this model, the counter-ions can move freely along the macromolecular "DNA subunit", the length of which is roughly defined by the persistence length of DNA. Since each subunit gives almost similar contribution to the polarizability of the macromolecule, the total polarizability divided by the DNA length should remain approximately constant. This result is valid even if the subunit length varies depending, e.g., on the DNA concentration [40].

The polarizability values found from the literature varies from ~$10^{-36}$ to ~$10^{-34}$ Fm$^2$/bp [22, 33, 40, 41]. For 2.7 kbp pUC18 plasmid DNA, Suzuki *et al* found the value ~$10^{-32}$ Fm$^2$ (~$4 \cdot 10^{-36}$ Fm$^2$/bp) [22]. For 12 kbp pTA250 plasmid DNA, Bakewell *et al* determined the polarizabilities as a function of frequency, yielding values from $0.14 \cdot 10^{-30}$ Fm$^2$ (~$2 \cdot 10^{-35}$ Fm$^2$/bp) for 5 MHz to $2.4 \cdot 10^{-30}$ Fm$^2$ (~$2 \cdot 10^{-34}$ Fm$^2$/bp) for 0.1 MHz [33, 41]. Saif *et al* has found the polarizability ~$5 \cdot 10^{-33}$ Fm$^2$ (~$6 \cdot 10^{-35}$ Fm$^2$/bp) for calf thymus DNA using a bit higher frequency (12.3 MHz) [42]. The polarizabilities per base pair we obtained in the case of relatively long DNA fragments (1065, 5141 and 8416 bp fragments in figures 5 and 6) correspond to the range of values found from the literature. Observed differences between the experimentally obtained polarizability values may also be caused by the use of different buffer, e.g., viscosity [23], or the length and the shape of DNA, e.g., the plasmid DNA has a circular conformation and also a globular shape secondary structure, which limits the unwinding and stretching of the plasmid DNA during DEP [22] and may result the weakening of the polarizability.

In contrast, in the case of short DNA fragments (27, 145 and 444 bp fragments in figures 5 and 6) it was observed that the polarizability per base pair does not remain constant but increases as a molecule get shorter. There are two likely reasons causing this kind of behaviour. First of all, short DNA fragments, i.e., which have a contour length of the order of "DNA subunit" length (about the persistence length of DNA ~150 bp), behave more like a rigid rod (in contrast to a globular ball in the case of long DNA), which enhances their polarizability in a longitudinal direction (a globular ball polarizes in many directions) [34]. Secondly, if the fragments are shorter or of the order of thickness of counter-ion cloud (~10 nm), the polarizability per base pair is enhanced by the counter-ion cloud polarization (discussed before in section 5.2). This enhancement of the polarizability has not been observed in the earlier studies [22, 33, 41, 42], because used DNA was long (2.7 - 12 kbp) compared to the fragments that were used in this study.

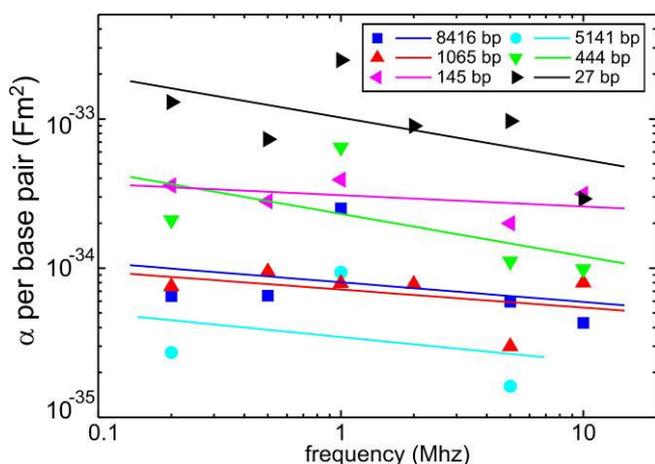

**Figure 5.** Experimentally obtained polarizabilities for the different size DNA fragments plotted as a function of frequency. The polarizability per base pair is obtained by dividing the polarizability of the whole molecule by its length in base pairs.

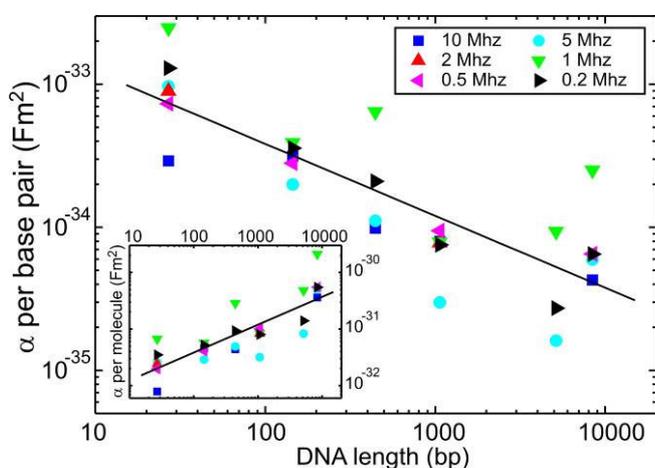

**Figure 6.** Experimentally obtained polarizabilities (per base pair) plotted as a function of the length of the DNA fragment with different frequencies. The polarizabilities for whole molecules are plotted in the inset.

## 6.   Immobilization of DNA

### 6.1.   DNA with and without thiols

Using the same methods we also compared the immobilization of C6-DNA, DTPA-DNA and (unmodified) 444 bp DNA to the finger-tip electrodes (See figure 7). The molecules without any thiol-linker diffused away from the DEP trap (fluorescence goes to zero) very soon after the trapping voltage was turned off as seen from the figure 7e. This indicates that the amount of a non-specific physisorption [43] is quite small. However, in the cases of both C6-DNA and DTPA-DNA the amount of the remained fluorescence was noticeably higher, indicating that the DNA was immobilized on the gold electrodes. Also, C6-DNA seems to attach better than DTPA-DNA.

After the trapping voltage is turned off (See figures 7a and c) the fluorescence intensity decreases almost linearly, faster than observed in the bleaching tests (See section 3.3). Thus, the negative slope is most likely due to diffusion of excess (e.g. because of the Coulomb repulsion between charged molecules) or poorly attached (weak physical adsorption of DNA to the gold surface) DNA away from the gap.

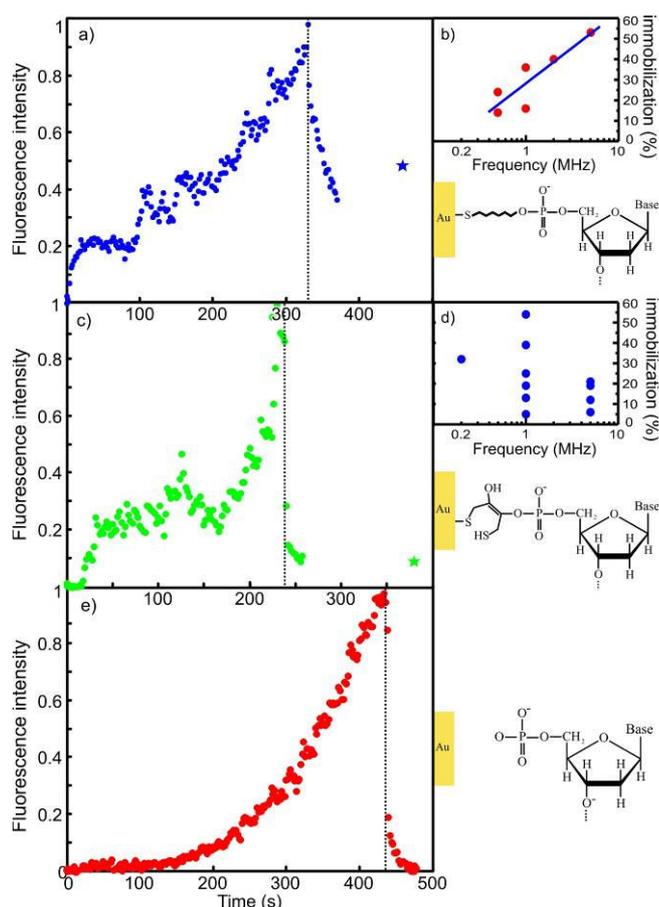

**Figure 7.** Trapping and immobilization of a) C6-DNA, c) DTPA-DNA and e) unmodified 444 bp DNA to the fingertip electrodes using DEP with 5 MHz ac signal (0.5 mM TCEP-HCl was used in the buffer). The circles describe the data points obtained from the DEP movie and the stars represent the 'remained fluorescence' measured separately after the trapping voltage had been off for a while. The dashed lines represent the times when the voltage was turned off. The relative immobilized amounts of b) C6-DNA and d) DTPA-DNA as a function of the signal frequency were obtained by comparing the remained amount of fluorescence with the maximum trapped fluorescence in each sample.

*6.2.   Effects of the frequency on immobilization of thiol-modified DNA*
The remained fluorescence divided by the maximum fluorescence observed during the DEP is plotted in figures 7b and d as a function of a frequency of the applied signal. C6-DNA seems to attach better by using the higher frequencies, which can be seen from figure 7b. This may be understood by the fast bonding of the highly reactive hexanethiol-linker together with the better localization of the DNA spot when using higher frequencies [25].

The immobilization results for DTPA-DNA seem quite inconsistent (See figure 7d) and have no observable regular behaviour. This is an indication of a poor binding of the linker, which may be due to the chemical structure of the DTPA linker, i.e., sulphur atoms in a S-S –form as a part of the ring. Before forming of an Au-S bond, the S-S bond must break, which could make it less favourable reaction than the bonding of –SH group of the hexanethiol-linker. The DTPA may physically attach [43] to the surface first and chemical binding may take place after a certain reaction time. This would result a slower time-scale of the immobilization, which may result to better attachment in the case of lower frequencies (which could be barely observed in figure 7d). It may also be that DTPA is bound to gold only through the physical adsorption.

*6.3.   Calculation of bonding energies*
Bonding of the hexanethiol and DTPA linkers on gold was studied also by the density functional theory (DFT) calculations [44]. The gold electrode was modelled by a tetrahedral twenty-atom gold cluster (See figure 8a). Vertex, edge and face-centered atoms in the cluster have three, six and nine nearest-neighbours, respectively, and provide convenient models for adsorption sites with various local chemical properties. We found that the optimal binding site for a hexanethiolate is a bridging position at the edge of the cluster (See figure 8b), with a binding energy of 1.8 eV. On the other hand, we could not find a stable adsorption site on the cluster for a DTPA linker where the S-S bond was left intact. Partial hydration of the S-S bond resulted in binding of the linker at the bridging edge position by 1.2 eV (See figure 8c and d). Although our calculations do not include effects from the environment, we note that the calculated binding energies agree qualitatively with the observed weaker attachment of DTPA-DNA as compared to the C6-DNA (See section 6.2).

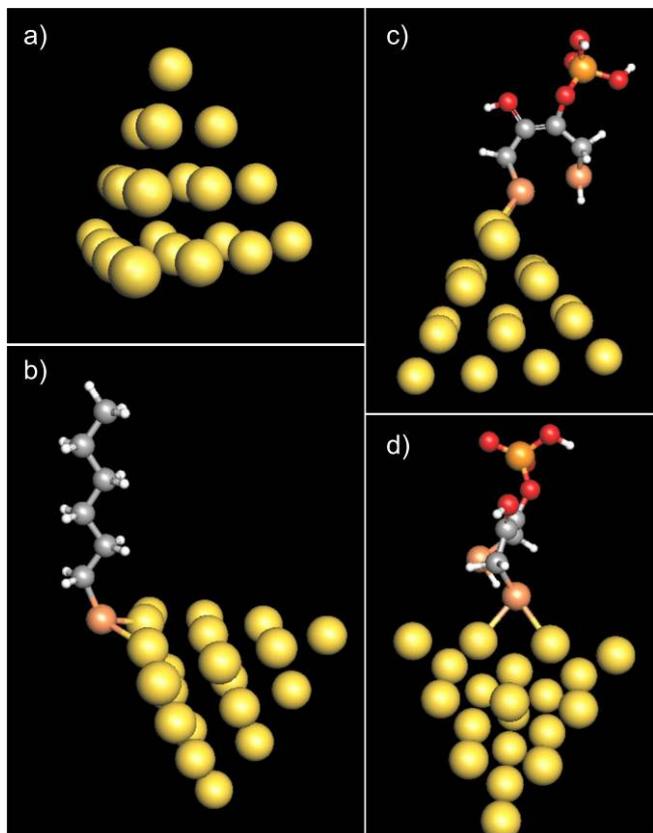

**Figure 8.** (a) The tetrahedral 20-atom gold cluster used for modelling of binding of the DNA linkers to gold, by the density functional theory calculations. Optimal binding configurations of (b) hexanethiol and (c, d) DTPA linkers on the cluster. (c) and (d) give two different views of the same configuration. The S-Au distances are: (b) 2.44 Å and 2.45 Å; (c) 2.45 Å and 2.51 Å. The S-S distance is 3.34 Å in (c).

## 7. Summary

The dielectrophoretic trapping of DNA fragments of different lengths, varying from 27 bp to 8 kbp, was studied *in situ* under a confocal microscope using frequencies from 0.2 to 10 MHz. Increase in the trapped amount of DNA inside the 'DEP trap', i.e., the constriction between fingertip type nanoelectrodes, as a function of the trapping voltage was recorded and carefully analyzed. Using the voltage and frequency dependent fluorescence data obtained from the experiments and utilizing finite-element electric field simulations, information about the polarizability of the DNA fragments was revealed. We obtained information on the field, frequency and DNA length dependence of DEP and DNA polarizability, as described below.

    Dependence of DEP on the electric field strength: It was observed that more voltage is needed to realize trapping in the case of smaller DNA fragments, which is due to the smaller polarizability and higher Brownian motion in the case of smaller molecules. It was also observed in the experiments, and explained by means of the electric field simulations, that small changes in the separation of nanoelectrodes, e.g., gap size of 80 nm compared to 130 nm, do not affect to the trapping efficiency as long as the separation is smaller than the physical dimension of the resolved "trapping region". Naturally, larger differences in separation of the electrodes can have a dramatic effect on trapping.

    Dependence of DEP and DNA polarizability on frequency: In general, higher trapping voltages were needed in the case of higher frequencies, which is probably due to the timescale of the counter-ion polarization. Keeping the trapping voltage fixed, more DNA was gathered with lower frequencies. On the

other hand for higher frequencies the DNA is better localized between the fingertip electrodes. This trade-off between efficiency and accuracy results into the optimum frequency which was found to be ~1 MHz.

Dependence of DNA polarizability on the length of DNA: Shorter DNA molecules had smaller polarizability than the longer ones. Interestingly, the polarizability *per base pair* was higher for small molecules than for large ones. This indicates that long DNA molecules do not behave as rod-shaped objects in the DEP process and that the polarization of DNA is related to the counter-ion cloud fluctuations.

Immobilization of two different types of thiol-linkers, namely hexanethiol and DTPA, on gold nanoelectrodes was demonstrated and compared with the non-specific binding of unmodified DNA fragments. The hexanethiol-linker was observed to attach better and more consistently than the DTPA linker, which was also suggested by our density functional calculations.

The DEP trapping and immobilization method we present here can be used as an efficient tool in the fabrication of molecular electronics circuits of different dimensions, from single molecules to complex DNA structures. In addition, the self-assembled devices build using DNA as a scaffold may be positioned in controlled way using DEP of DNA.

**Acknowledgement.** This work, conducted as part of the award (Nanoscale quantum systems interacting with fields: ultracold gases and molecular electronics)  made under the European Heads of Research Councils and European Science Foundation EURYI (European Young Investigator) Awards scheme, was supported by funds from the Participating Organisations of EURYI and the EC Sixth Framework Programme. We also acknowledge the financial support of Academy of Finland (project number 205470), the National Graduate School in Informational and Structural Biology, and the National Graduate School in NanoScience.